\documentclass{article}
\usepackage{microtype}
\usepackage{graphicx}
\usepackage{booktabs}
\usepackage{array}
\usepackage{multirow}
\usepackage{amsmath,amssymb}
\usepackage{float}
\usepackage{algorithm}
\usepackage{algorithmic}
\usepackage{enumitem}
\usepackage{capt-of}
\usepackage{cuted}
\usepackage{flafter}
\usepackage[hyperfootnotes=false]{hyperref}
\usepackage{xurl}

\usepackage[accepted]{mlsys2025}
\usepackage{comment}
\usepackage{etoolbox}
\newcommand{\sys}{Ripple\xspace}
\newcommand{\atom}{unit\xspace}
\newcommand{\atoms}{units\xspace}
\newcommand{\budget}{budget\xspace}
\newcommand{\budgets}{budgets\xspace}

\newcommand{\supp}[1]{}
\newlength{\eightlinefigureheight}
\AtBeginDocument{\setlength{\eightlinefigureheight}{8\baselineskip}}
\newcommand{\outlinepoint}[1]{\paragraph{#1}}

\newcommand{\workloadmix}[3]{\ensuremath{(Robotic_{#1},\,LLM_{#2},\,VLM_{#3})}}

\input{paper_config}
\usepackage{colortbl}
\definecolor{tok}{HTML}{CDEBD7}
\definecolor{toff}{HTML}{F6C9C6}
\definecolor{tred}{HTML}{D62728}

\makeatletter
\renewcommand{\Notice@String}{}
\patchcmd{\printAffiliationsAndNotice}
  {{\bf AUTHORERR: Missing \textbackslash{}mlsyscorrespondingauthor.}}
  {}{}{\PackageError{tide}{Could not suppress the missing-contact warning}{Check the MLSys style definition.}}
\makeatother
\mlsystitlerunning{Ripple: Deadline-Aware CPU--GPU Sharing for Embodied Inference}
\hypersetup{pdftitle={Ripple: Deadline-Aware CPU--GPU Sharing for Embodied Inference},
  pdfsubject={CPU--GPU memory bandwidth sharing on edge devices}}

\begin{document}
\twocolumn[
\mlsystitle{Ripple: Deadline-Aware CPU--GPU Sharing for\texorpdfstring{\\}{ }%
Embodied Inference}
\begin{mlsysauthorlist}
\mlsysauthor{Jiadong Zhang}{mbzuai}
\mlsysauthor{Xiaosong Ma}{mbzuai}
\end{mlsysauthorlist}
\mlsysaffiliation{mbzuai}{MBZUAI}
\mlsyskeywords{LLM serving, unified memory, admission control, deadline scheduling}
\vskip 0.3in
\begin{abstract}

Unified-memory edge devices increasingly run LLM/VLM inference alongside time-critical CPU tasks for perception and control.
GPU memory traffic can overwhelm these tasks, yet platforms such as DGX Spark expose neither bandwidth counters nor direct bandwidth regulation.
We present \sys{}, a CPU--GPU co-scheduler that combines model-derived GPU memory volumes with CPU completion feedback to regulate interference.
\sys{} divides inference into pausable, layer-based units and translates each CPU job's deadline slack into a GPU-byte budget, admitting work within the budgets of all overlapping jobs.
Online updates track changing interference, while deadline-aware pacing and judicious SM footprint narrowing preserve inference progress.
Implemented with SGLang engine hooks and a central scheduler, we evaluate \sys{} on DGX Spark with four robotic CPU task mixes and LLM/VLM inference.
It is the only evaluated approach that consistently meets the 2\% CPU deadline-miss target and VLM deadlines across the tested deadline settings while retaining best-effort LLM throughput.
\end{abstract}

]
\printAffiliationsAndNotice{}

\section{Introduction\pagebudget{1.75}}
\label{sec:intro}

AI models increasingly run on edge devices~\cite{applem1,sparkoverview},
with privacy protection, low/consistent latency, low cost, and device portability.
Many of these devices adopt a unified memory architecture (UMA), where the CPU and GPU share one physical memory on one package, as with Apple's M-series processors and NVIDIA's DGX Spark. 
UMA is cheaper, smaller, and easier to build and maintain, eliminating host-device memory copies and providing a decent capacity (128GB on DGX Spark) to hold reasonably large models (\eg, 32B at FP16).
With the rapid growth of agentic and embodied AI, edge AI devices are increasingly \emph{versatile}, often pairing model inference with CPU computation tasks for perception and control, which typically come with timing requirements.
Consider a home robot working on clearing a dinner table while interacting with human users on travel planning. 
Every few 10s of milliseconds its CPU must work out how far away things are and where it can safely move, or the robot may walk
into a person who has just stepped into its path; 
every few seconds a vision-language model (VLM) examines the table again; 
meanwhile a large language model (LLM) plans the next robot movements, retrieves family travel history/preference and calendar events using retrieval-augmented generation (RAG)~\cite{lewis2020rag}, and generates responses to the user. 
A missed obstacle-avoidance cycle can cause an accident, a
late scene update stalls the next action, and a slower answer merely arrives later --- all three run on one chip and share its memory bus~\cite{nav2marathon,palme,sayplan,partnr}.

\begin{figure}[!tbp]
\centering
\IfFileExists{figures/00-foreclosures.pdf}{%
\includegraphics[width=.8\columnwidth]{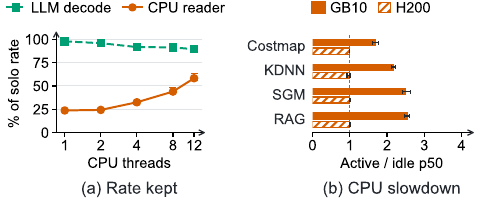}%
}{%
  \fbox{\parbox[c][\eightlinefigureheight][c]{0.93\columnwidth}{\centering\small
  \textbf{Figure pending:} \texttt{figures/00-foreclosures.svg}\\[2pt]
  Layout placeholder, no data drawn.\\
  Brief in \texttt{figures/svg-prompts.md}, F00.}}%
}
\caption{CPU--GPU interference under LLM decode.}
\label{fig:foreclosures}
\label{fig:intro-harm}
\end{figure}

Our study finds that \emph{memory bandwidth is the contention point, in a lopsided fight,} dominated by the less time-critical inference tasks (user response generation in the above scenario).
Figure~\ref{fig:intro-harm} showes this with a workload mix on the GB10 processor of DGX Spark, with 4 routine robotic/agentic CPU tasks
--- costmap updates, nearest-neighbor queries, stereo matching, and vector retrieval over the robot's memory. 
The left chart reports the mutual slowdown when this CPU mix co-runs with a GPU-side LLM decode task, with growing CPU thread allocations (CPU-side performance saturating at 12). 
The more deadline-sensitive CPU tasks suffer much higher loss. 
The right chart provides a reference by running the same test on an H200 server, with separate HBM. 
The CPU tasks sees no impact of concurrent LLM inference there, as opposed to 1.7-2.5$\times$ on the Spark.
\noindent Managing bandwidth on these devices is harder than on a conventional CPU, where we have homogeneous cores and a rich set of hardware counters.
On edge AI devices with UMA, bandwidth management faces \textbf{\emph{four major challenges}}: 
(1) the platform lacks bandwidth counters and exposes no interface that reports per-requester memory bandwidth, neither that regulates bandwidth consumption;
(2) modern transformer-based AI models are bandwidth-intensive even at token granularity: on GB10 one LLM token has a 69-74ms forward pass and one VLM image a 285ms one, longer than the 45 to 70ms deadlines for recurrent CPU robotic tasks;
(3) the CPU side is not designed to take most of the unified memory bandwidth, consuming at most 60\% of it even with all cores;
(4) the power-aware design of edge devices perform aggressive and black-box, temperature-triggered clock-throttling, further complicating performance monitoring and resource regulation.

These constraints rule out most existing (and relatively mature) multi-tenant bandwidth management solutions~\cite{memguard,bwlockpp,rdt2022,libsmctrl,nvtaskset,conserve,migguide}, because the interfaces they rely on are missing.
The one scheduler to our knowledge that work without, published recently, is \textsc{Sereno}~\cite{sereno}, which protects foreground
mobile apps from background NPU inference. 
It targets a phone SoC, where the bandwidth relinquished by the NPU can be almost entirely absorbed by the CPU.
Edge UMA devices are not zero-sum in this way.
Also, \textsc{Sereno} pairs one foreground app with one background model, where our setting has many concurrent workloads on both sides, each with its own timing
requirement.

%

This work is based on the observation that ``what the CPU and GPU cannot do symmetrically, they can do complementarily.'' 
GPU inference is \textit{opaque in time and transparent in volume}: when a forward pass will finish depends on contention and thermal throttling, but its memory access volume is well known.
A real-time CPU task is the exact opposite ---
\textit{transparent in time and opaque in volume}: 
its release time and deadlines are set in stone, whether each deadline is met can be easily known,
but access volume remains not observable without detailed source code instrumentation.
Together, the two pieces of information give 
sufficient guidance for our scheduling, compensating for the lack of bandwidth counters (or even surpassing their capability).

To this end, we propose \sys, a CPU-GPU co-scheduling framework that performs fine-grained GPU inference task admission, to protect CPU-side deadline-sensitive tasks while retaining overall bandwidth utilization and inference speed as allowed by the deadlines. 
\sys overcomes the lack of counters by leveraging the semantic knowledge on today's mainstream transformer-based LLM/VLM workflow to decompose the inference forward pass into smaller, layer-based work units with known memory access volumes, for GPU-side bandwidth regulation.  
It performs continuous modeling by observing the GPU-side memory traffic on CPU-side deadline-sensitive tasks, leading to a pricing model that releases metered volume budgets to admit GPU work units, dynamically adjusted to current clock throttling levels and CPU-side contentions.

We implement \sys{} as a standalone scheduler with engine hooks in SGLang and evaluate it on DGX Spark with four mixes of robotic CPU tasks plus LLM/VLM inference.
It is shown to consistently meet CPU deadline-miss targets and VLM deadlines, while retaining substantial best-effort LLM throughput across workload mixes and deadline settings.

\para{Novelty}
To the best of our knowledge, we are the first to target the bandwidth asymmetry on DGX Spark.
Also, \sys{} is the first system that makes independent CPU tasks' declared deadlines and miss tolerances explicit constraints for LLM/VLM co-serving on UMA devices.
It introduces the first counter-free scheduler that translates CPU deadline slacks into GPU memory-byte budgets,
with the first theoretical throughput bound for best-effort inference under these CPU timing constraints.

\para{Scope.}
\label{sec:discussion}
This proof-of-concept study targets declared deadline miss tolerances and batch-one, pausable inference.

\section{Background and Motivation}
\label{sec:motivation}
\subsection{Unified Memory at the Edge}
\label{sec:pool}
\label{subsec:uma}
\begin{table}[!t]
\caption{Memory and %
power of representative GPUs.
``--'' denotes missing vendor specifications~\cite{applem4max,sparkhardware}.}
\centering
\setlength{\tabcolsep}{1.5pt}
\small
\begin{tabular}{@{}llrrrrc@{}}
\toprule
Device & Memory & Size & BW & BW & Power & Shared \\
 & & (GB) & (GB/s) & /GB & /GB & w/ CPU \\
\midrule
\multicolumn{7}{@{}l}{\textbf{\emph{Edge, UMA}}} \\
M4 Max & LPDDR5 & 128 & 546 & 4.3 & -- & yes \\
\shortstack[l]{Ryzen AI\\Max+} & LPDDR5X & 128 & 256 & 2.0 & 0.9 & yes \\
Jetson Orin & LPDDR5 & 64 & 205 & 3.2 & 0.9 & yes \\
Jetson Thor & LPDDR5X & 128 & 273 & 2.1 & 1.0 & yes \\
DGX Spark & LPDDR5X & 128 & 273 & 2.1 & 1.1 & yes \\
\midrule
\multicolumn{7}{@{}l}{\textbf{\emph{Discrete}}} \\
RTX 5090 & GDDR7 & 32 & 1792 & 56 & 18 & no \\
\midrule
\multicolumn{7}{@{}l}{\textbf{\emph{Datacenter}}} \\
H100 NVL & HBM3 & 94 & 3900 & 41 & 4.3 & no \\
H200 NVL & HBM3e & 141 & 4800 & 34 & 4.3 & no \\
MI300A & HBM3 & 128 & 5300 & 41 & 4.3 & yes \\
\shortstack[l]{GB300\textsuperscript{a}} & HBM3e & 288 & 8000 & 28 & 4.9 & no \\
\bottomrule
\end{tabular}
\par\medskip
\begin{minipage}{\columnwidth}\small
\textsuperscript{a}{\small ~Per GPU of GB300 NVL72. Its Grace CPU has its own LPDDR5X~\cite{blackwellultra,gb300}.}
\end{minipage}

\label{tab:uma}
\end{table}

\textit{Unified Memory Architecture (UMA)}~\cite{glaskowsky1999} began as a power- and cost-driven solution in integrated graphics~\cite{intel810dvmt,llano}, enabling a small GPU to ``borrow'' system memory instead of carrying its own. 
It has since become the common design for AI-capable edge devices: Apple's M-series processors, AMD's Ryzen AI Max, NVIDIA's Jetson modules as well as DGX Spark all come with a single memory, of up to 128GB, 
directly accessed by both the GPU and CPU side. 
Table~\ref{tab:uma} summarizes these recent models and their major parameters~\cite{igpuhistory,applem1,applem4max,amdryzenaimax,jetsonorin,jetsonthor,sparkoverview}.

All the above devices with UMA adopt \textit{LPDDR (Low-Power Double-Data-Rate memory)}~\cite{jedeclpddr5}. 
Unlike the DDR modules of desktops and servers, LPDDR is designed for battery-powered devices: it runs at lower voltages, enters deep power-saving states when idle, and is soldered next to the processor rather than socketed.
It offers large capacities at low power and price, with modest
bandwidth, as reflected by the bandwidth/GB column and power/GB columns. 
Edge GPUs therefore raise their bandwidth by widening the LPDDR bus, to 256 bits on DGX Spark, or by raising its pin rate, as Jetson Thor does over Orin~\cite{jedeclpddr5,jedechbm3,jedecgddr7}.
Such limited bandwidth is further shared between GPU and CPU.

This stands in stark contrast to datacenter GPUs, also shown in the table, where the more familiar HBM (High Bandwidth Memory) delivers 
13 to 20$\times$ the bandwidth/GB of DGX Spark, with rated chip/module power of 4.3--4.9\,W per GB of memory. 
Among the datacenter models listed, only the AMD MI300A shares its memory between GPU and CPU, and its bandwidth/GB is around 20$\times$ that of DGX Spark.
Discrete accelerators, such as the NVIDIA RTX 5090, adopt GDDR (Graphics DDR) memory, which drives each pin several times faster at higher power and delivers a bandwidth/GB even higher than the datacenter GPUs
(Table~\ref{tab:uma};~\citealp{rtxblackwell,mi300aupm,blackwellultra,h100spec,h200spec,mi300a,mi300aoverview}).
%

%

%

\subsection{Memory Bandwidth Contention on DGX Spark}
\label{sec:contention}
\label{sec:determinants}
\label{sec:memory-sharing}
\label{sec:memory-attribution}
\label{sec:harm}
\label{sec:background}
\label{subsec:contention}
For the rest of the paper, we focus our discussion on the NVIDIA DGX Spark architecture (``GB10'' for short). 
Released in October 2025, it has rapidly grown into a product line sold by seven major PC makers and become a mainstream edge AI device~\cite{sparklaunch}.
Neither the bandwidth contention problem on UMA devices (Table~\ref{tab:uma}) nor our proposed approach, meanwhile,
is specific to GB10.
\begin{figure}[!htb]
\centering
\includegraphics[width=.8\columnwidth]{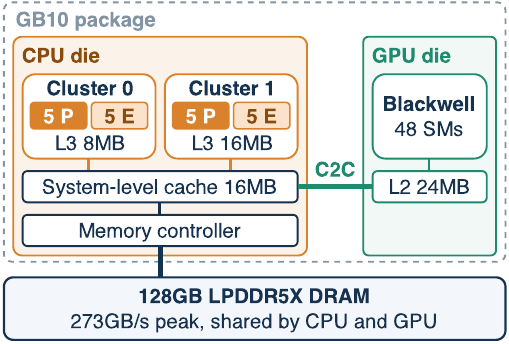}
\caption{GB10 package and memory topology
}
\label{fig:topology}
\end{figure}

\para{DGX architecture}
GB10, the DGX Spark processor, places a CPU and a GPU die on a single package (Figure~\ref{fig:topology}). 
The CPU has 20 cores in two clusters, each containing 5 performance (P) and 5 efficiency (E) cores.
The 48-SM GPU die reaches the 128GB shared LPDDR5X DRAM through NVLink-C2C and the memory controllers on the CPU die. 
Our benchmarking achieves 190-257GB/s combined DRAM throughput out of the 273GB/s from NVIDIA specification, depending on the access pattern (Appendix~\ref{app:channel}).

%

\begin{figure}[!htb]
\centering
\includegraphics[width=.9\columnwidth]{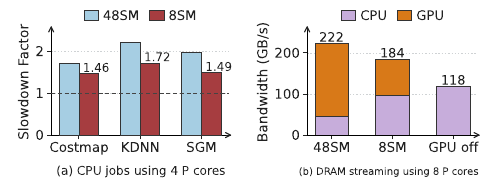}
\caption{(a) Slowdown brought by GPU inference (with different SM counts) on CPU time-sensitive tasks;
(b) bandwidth with concurrent CPU+GPU memory streaming
}
\label{fig:levers}
\label{fig:cores}
\label{fig:sm}
\end{figure}

%
%
\label{sec:why}
Sharing introduces contention, especially with the low overall bandwidth availability relative to capacity.
GB10 provides no direct bandwidth allocation mechanism between its CPU and GPU sides:
how bandwidth divides appears to depend on how many memory requests each keeps in flight.
This gives the GPU side an easy advantage, where thousands of memory accesses can be issued with its massive parallelism, burying the small number of time-critical and latency-sensitive CPU requests.
The above asymmetry is too strong to mitigate through high-level concurrency parameters, in other words, CPU core or GPU SM counts, as shown earlier in Figure~\ref{fig:levers}.
%

%

%
%
%

%
%
%

%
%
%
%
%
%
%
%

\para{Bandwidth Management Challenges}
\label{sec:constraints}
\label{sec:controls}
\label{sec:interfaces}
\label{sec:cpu-side}
\label{sec:existing}
In addition to such significant CPU-GPU imbalance, \agreed{UMA devices like GB10 pose} unique challenges for bandwidth management: %

\par\noindent\textbf{\textit{No observation or regulation}} 
Unlike server CPUs, with mature per-workload memory bandwidth monitoring/allocation mechanisms such as Intel RDT or ARM MPAM~\cite{rdt2022,armmpam},
the GB10 firmware exposes no memory-controller counters and does not enable MPAM. 
Our tests found that NVIDIA DCGM's DRAM-activity metric returns no reading.
\par\noindent\textbf{\textit{Limited GPU control}}
An MPS SM limit is set when a process starts and remains throughout its execution.
Lowering the GPU clock reduces GPU application throughput while freeing little bandwidth.
The recent CUDA \emph{green contexts}~\cite{nvidiagreenctx} can confine a stream's kernels to a subset of the SMs. 
Our measurement (Figure~\ref{fig:levers}(a)), however, demonstrates that reducing LLM's GPU SM count from the full 48 to 8 (minimum allowed) has very limited effect on mitigating CPU slowdown.
Meanwhile, with the GB10 CPU clusters unable to saturate the memory bus, when both the GPU and CPU sides run memory streaming benchmarks, the bandwidth yielded by reducing the GPU SM count cannot be absorbed by the CPU, resulting in low overall bandwidth utilization, as shown in Figure~\ref{fig:levers}(b). 

\par\noindent\textbf{\textit{Run-to-run performance variability}} 
Being power- and heat-aware, ddge devices constantly adjust CPU/GPU clock rate to temperature, further complicating bandwidth control.
\subsection{Major Edge Workloads}
\label{sec:workloads-bg}
\label{sec:d2}
\label{sec:d3}
\label{sec:memory-timing}
\para{CPU-side}
Representative CPU tasks on edge AI devices, in the current pursuit of AGI, are diverse and often tightly timed.
Embodied systems run their periodic perception, localization, and control loops on CPU: constructing LiDAR-based occupancy grids, stereo disparity and image features, nearest-neighbour queries for scan matching, and model-predictive control.
These loops run with fixed frequency, typically at 10-50Hz with deadlines of tens of milliseconds.
Agentic systems retrieve from their memory before responding: in a spoken conversation that retrieval must return within the few hundred milliseconds of a turn.
\begin{table}[!htb]
\begingroup
\caption{Common edge robotic CPU tasks~\cite{kdtree,sgm,stivers2009,orb,kitti,nav2marathon,osqp,generativeagents,nav2controller}.}
\par\vspace{0.1in}
\label{tab:workloads}
\centering
\small
\setlength{\tabcolsep}{3pt}
\renewcommand{\arraystretch}{1.15}
\begin{tabular}{@{}l>{\raggedright\arraybackslash}p{0.76\columnwidth}@{}}
\toprule
Task & Description \\
\midrule
Costmap & Maps nearby obstacles from LiDAR scans. \\
KDNN & Finds nearby points to help align scans and track position. \\
SGM & Estimates object distances from two camera images. \\
Control & Computes quadrotor control commands using model-predictive control (OSQP). \\
ORB & Extracts visual features from camera images. \\
\bottomrule
\end{tabular}
\endgroup
\end{table}

\agreed{Table~\ref{tab:workloads} lists representative CPU tasks in embodied AI workloads.} 
Unlike GPU tasks, they tend to be complex yet resource-light.
Each of the above tasks has its algorithm and access pattern, but moves at most 1.6GB of data per iteration, about 1/10 of what an LLM moves in generating one token (\eg, 15.2GB for Qwen3-8B).
A task typically tolerates a small fraction of misses, as in the weakly-hard model~\citep{weaklyhard} of real-time systems.
The tasks run with independent intervals and phases, with overlapping deadline windows.
\para{GPU-side}
GPU workloads on edge AI devices, on the other hand, are computationally intensive, with rather elastic timing requirements.
Examples include LLMs for planning and answering and VLMs (Vision-Language Models) for scene understanding, reasoning, and task planning in embodied robots.
With all mainstream models, a forward pass goes through a regular sequence of layers, with per-layer model weights streamed from DRAM.
Edge LLM/VLM inference processing is particularly bandwidth-hungry during decode, where a small batch (mostly 1) reads GBs per token.
The small batch sizes, limited model sizes, and limited context lengths in edge AI scenarios also make KV cache access much lighter than in datacenter services:
our 8192-token context used with Qwen3-32B-FP8 yields KV cache memory access at around 6.5\% of model weight read volume.

Between LLM and VLM, 
the latter requires timely updates to its understanding of the environment to support reasoning and task planning, \agreed{though with an interval often much longer than CPU-side perception/control tasks.}
Also, they
spend host time differently: a VLM decode step issues about a thousand kernels lasting 28\,$\mu$s on average, making CUDA graph replay indispensable to avoid a host bottleneck;
the LLM kernels are fewer and last much longer, 
receiving negligible performance gain from CUDA graphs.
Finally, running the LLM and VLM engines on GB10 side by side has little GPU SM-level interference: DRAM bandwidth is the clear contention point (Appendix~\ref{app:bytes}).

\section{Approach Overview\pagebudget{1}}
\label{sec:overview}
\label{sec:challenges}
\label{sec:requirements}
\para{Rationale}
In this work, we leverage several opportunities to overcome the lack of UMA bandwidth monitoring or regulating mechanisms. 
First, the time-critical CPU tasks do not require full dedication of CPU resources or DRAM bandwidth, so 
uniform GPU throttling would be an overkill. 
Second, the many layers of LLM/neural-network models provide a natural granularity of GPU-side task admission, with known data volume, to allow bandwidth regulation without wasting resources on aborted work.
Finally, one does not need a detailed hardware counter for our purpose: the coarse observation based on the completion times of past jobs, plus the GPU-side DRAM access volume during each deadline window, provide sufficient information for task-level coordination. 
Note that a task's completion time already reflects how much the job itself read and how fast the clocks ran, giving a complete CPU-side task profile.
%

%
%

Based on these observations, we present \sys, a CPU-GPU task scheduler for UMA devices that protects time-critical CPU tasks while retaining the progress of bandwidth-hungry GPU inference. Most importantly, it does so without requiring a bandwidth counter/arbitrator.
Instead, it focuses on informed GPU task admission, predicting the delay the task adds to every active CPU job.

Note that GPU work could also be deadline-bound, such as VLM inference for the scene updates that guide the robot's next action.
However, due to the biased memory arbitration discussed in Section~\ref{sec:contention}, only the CPU tasks need \textit{bandwidth protection}: deadline-bound GPU tasks are safe as long as \sys gives them scheduling priority.  
\para{System model and objective}
Following the above reasoning, we term the time-critical CPU workloads \emph{protected tasks (PTs)}. 
Each $PT_i$ releases a sequence of jobs and is associated with an inter-arrival interval $T_i$: known and fixed for periodic tasks such as costmap updates and stereo matching, and a declared minimum for sporadic tasks such as memory retrieval for a spoken reply. 
$PT_i$ also comes with a relative deadline
$D_i \le T_i$ and a tolerated deadline miss rate (DMR) $\varepsilon_i$. %
Every job reports its completion; \sys neither modifies nor throttles the task.

Each GPU inference workload is either \emph{deadline-bound} (with a per-request deadline), or \emph{best-effort}.
\sys observes only the release/completion time of a CPU job and the DRAM access volume and duration of each GPU job.
It learns about the CPU tasks online from their completions and requires little offline profiling.
The more predictable GPU inference tasks are profiled once per device and model, recording each \atom's completion time, each engine's overhead per grant, and the peak byte rate $\rho$.

In priority order, its objectives are
(1) that in every window of $n_i$ consecutive jobs of a protected task, at most a fraction $\varepsilon_i$ miss $D_i$, except with probability at most $\delta$ (a chance constraint~\cite{charnes1959}),
where $PT_{i,j}$ misses when its completion time $CT_{i,j}$, measured from its release $R_{i,j}$, exceeds $D_i$; 
(2) that every deadline-bound GPU workload meets its deadline; 
and (3) that best-effort GPU work makes as much
progress as possible. 
\sys acts only on the GPU work it schedules: when each piece is dispatched and on how many SMs. 
Non-time-critical CPU tasks or
the division of service among several best-effort
workloads (CPU or GPU) is beyond the scope of this work.

\para{Main idea}
\sys leverages the common software architecture for GPU model inference today for scheduling GPU work at an appropriate granularity. 
It cuts each engine's forward pass into \emph{\atoms}, the units at which execution can pause with state information in memory, and authorizes a contiguous run of \atoms, called a \emph{grant}, at a time. 

Each CPU job receives an \emph{allowance} $A_{i,j}$, the total GPU memory access volume (``volume'' for short) assigned from its predicted deadline slack. Its \emph{\budget} $B_{i,j}(t)$ is the volume available for additional GPU work at time $t$: the allowance released by then, less the volume already charged.
When an engine finishes a grant, it reports its execution time. The scheduler then takes the global \budget $B(t)$, the minimum across active uncertified CPU jobs, selects the longest grant that fits it,
and charges its volume to every active CPU job. 
When a CPU job completes, the scheduler updates that
task's model of its perceived interference from GPU volumes.
When the same PT's next job is released, \sys adjusts that job's allowance from the updated model.

\begin{figure}[H]
\centering
\includegraphics[width=.9\columnwidth]{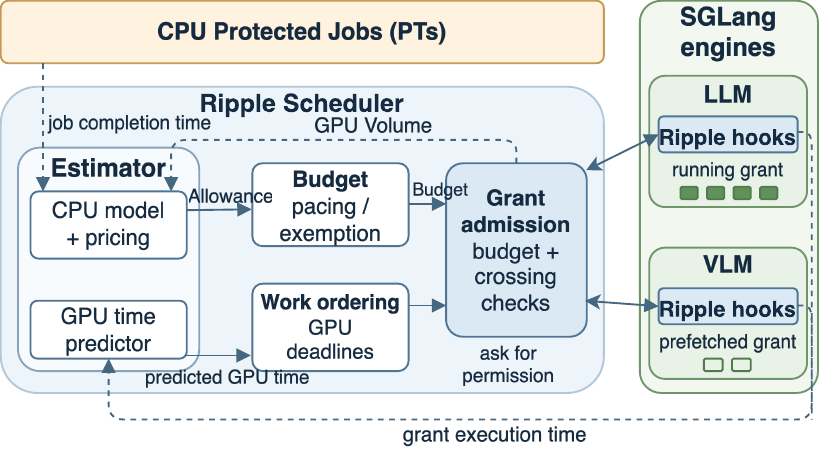}
\caption{\sys{}'s modules and their interactions, with solid lines indicating control flow while dashed ones feedback} 
\label{fig:structure}
\end{figure}

\para{Software architecture}
\sys comprises three major modules, as illustrated in Figure~\ref{fig:structure}. 
It adds \emph{engine hooks} in SGLang~\cite{sglang} that divide forward passes at \atom boundaries, pause engines between grants while preserving state, apply SM allocations through CUDA green contexts, and report each grant's CUDA time.
Its \emph{scheduler} runs as a separate process occupying one performance core; it maintains per-job byte accounts, orders GPU work, and admits grants within the CPU jobs' budgets.
It controls the SGLang engines (also running on CPU cores), one per inference service, via the \sys hooks. 
The \sys \emph{estimator} runs within the scheduler, updating each CPU task's model on job completion (dashed feedback edges) and computing allowances for subsequent jobs.
\section{\sys Design\pagebudget{2}}
\label{sec:design}
\label{sec:opportunities}

\subsection{Software Architecture and Scheduling Workflow}
\label{subsec:workflow}
\label{sec:interruptible-graphs}
\label{sec:vlm-pacing}
\sys manages UMA bandwidth through GPU-side task admission, with no interference with the scheduling of CPU-side tasks.
It deploys a central scheduler, along with its scheduling-relevant components through SGLang modification.
Such a \sys-customized SGLang scheduler performs bandwidth-aware LLM/VLM task decomposition/regrouping at layer boundaries, as well as the GPU task monitoring/reporting duties required by \sys.
The \sys scheduler, which makes all admission decisions, occupies one GB10 performance core by default, due to its heavy processing and time-sensitive task nature; the SGLang processes of the inference engines share a separate set of cores, disjoint from those of the PTs.

The chief goal of the \sys scheduler is to maximize overall DRAM bandwidth utilization (and consequently the GPU inference throughput), while keeping each time-sensitive task's deadline misses within its tolerance: PTs on CPU and deadline-bound GPU tasks.
It makes admission decisions at \emph{scheduling points} triggered by the following events:
(1) a CPU PT job starts or completes;
(2) a GPU grant completes;
(3) an engine announces new work; or
(4) the \budget accrues enough to schedule the next \atom.

On the CPU side, as mentioned earlier, \sys does not interfere with CPU tasks' scheduling or execution, but observes the slowdown they experience from GPU memory bandwidth usage (major) and other CPU tasks or environment variances such as thermal throttling (minor). 
\sys builds a per-PT interference model through linear regression and updates it constantly by new data points collected upon PT job completion. 
Based on the model, it assigns each PT job a total volume \emph{allowance} $A_{i,j}$ for its lifetime. 
At time $t$, $PT_{i,j}$'s \emph{budget} $B_{i,j}(t)$ is the released part of $A_{i,j}$ minus the GPU volume already charged to the job.
Section~\ref{subsec:pricing} details the slowdown modeling and allowance/budget calculation.

On the GPU side, \sys breaks an LLM/VLM forward pass into \atoms at the natural boundaries of modern transformer-based models: transformer layers, the output head, and the attention/MLP blocks of a vision encoder.
At each scheduling point, if possible, \sys admits the next \emph{grant}, the longest stretch of one engine's \atoms, that fits the global \budget $B(t)$, the minimum over active uncertified PT jobs.
Deadline-bound and best-effort \atoms are ordered by the \sys scheduler by their own deadlines: a deadline-bound request's work gets due times spread evenly over a fraction $\gamma$ of the interval to its deadline, and a best-effort grant runs before the next due time only if it ends by then and leaves enough \budget for the request's next grant.

As mentioned in Section~\ref{sec:workloads-bg}, of the two engines, only the VLM gains from CUDA-graph replay, so \sys keeps SGLang's breakable graphs and pauses between their segments.
The LLM, in contrast, runs eagerly and can leave and re-enter its forward pass at any \atom boundary.
A grant that an engine has started runs to its end, since work submitted to the GPU cannot be recalled.

\subsection{Pricing GPU Work Against CPU Deadlines}
\label{sec:cpu-admission}
\label{sec:interference-estimation}
\label{sec:recovery}
\label{sec:mispredict}
\label{subsec:pricing}

\sys{} models how GPU volume delays $PT_i$'s $j$th job:
\begin{equation}
CT_{i,j} = a_i + \kappa_i C_{i,j} + e_{i,j},
\label{eq:victim}
\end{equation}
where $a_i$ estimates the job's interference-free duration, $\kappa_i$ its delay by GPU volume, and $e_{i,j}$ the residual execution time not explained by the model.
Upon the job's completion, \sys refits $a_i$ and $\kappa_i$ by least squares with exponential forgetting~\cite{ljung1983}, following changes in measured completion times.
Using a conservative CPU duration estimate $\bar a_{i,j}$ and slope estimate $\kappa^u_{i,j}$, 
$PT_{i,j}$ has the nominal allowance
\begin{equation}
A_{i,j} = \min\Bigl(\frac{\max(0,\ D_i - \bar a_{i,j})}{\kappa^u_{i,j}},\ A^{\max}_i\Bigr),
\label{eq:allowance}
\end{equation}
converting the remaining deadline slack into allowed GPU volume.
Here $\bar a_{i,j}$ reserves time for the job's own execution, CPU-side inter-task interference (very minor), thermal throttling, and run-to-run variances.\footnote{Details of \sys's pricing model and budget management are given in Appendix~\ref{app:pricing}.}
$A^{\max}_i = \rho D_i$ caps tasks with $\kappa$ near zero (insensitive to bandwidth contention), where $\rho$ is the peak GPU memory access rate. 

$A_{i,j}$ is the total GPU volume allowed by job $PT_{i,j}$, 
released linearly over its predicted interference-free duration $\hat a_{i,j}$. 
At time $t$ within this interval following $PT_{i,j}$'s release time $R_{i,j}$, the job has a remaining \budget
\begin{equation}
B_{i,j}(t)=\frac{A_{i,j}\,(t-R_{i,j})}{\hat a_{i,j}}-C_{i,j}(t),
\label{eq:job-budget}
\end{equation}
where $C_{i,j}(t)$ is the volume already charged to the job. %

At a scheduling point (time $t$), \sys identifies the next grant, the maximum sequence of \atoms whose combined volume fits under the global \budget
\begin{equation}
B(t)=\min B_{i,j}(t), \text{for all active $PT_{i,j}$} 
\label{eq:global-budget}
\end{equation}
so that the volume is allowed by the most sensitive PT. 
Once issued, the GPU grant is charged in full to \emph{every PT it overlaps with}, as it simultaneously delay all of them.

\begin{figure}[H]
\centering
\includegraphics[width=\columnwidth]{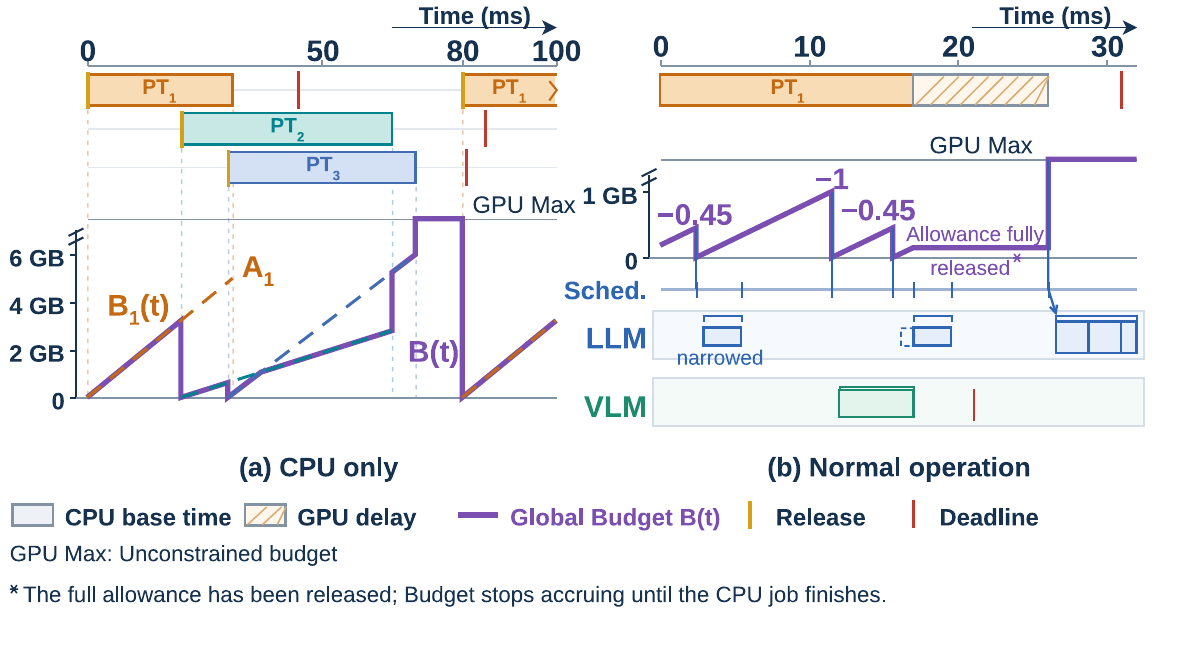}
\caption{\sys{}'s control loop illustrated with sample budget calculation and grant issuing
}
\label{fig:control-loop}
\end{figure}

\agreed{Figure~\ref{fig:control-loop}(a) illustrates the allowance/budget calculation with 3 PTs and no active GPU workloads.
The purple solid line marks the global budget $B(t)$, which follows the lowest active per-PT budget line (dashed lines).
Figure~\ref{fig:control-loop}(b) simplifies with only one PT, but adding two LLM/VLM inference tasks, with grants admitted subject to $B(t)$ (the illustrated LLM and VLM grants require 0.45\,GB and 1\,GB, respectively).
When $PT_1$ completes, \sys is able to release more LLM \atoms in a single grant. 
Different height of the grant boxes indicate SM narrowing applied only to LLM decode.
}

%
%

\subsection{Dynamic SM Regulation and Overhead Mitigation}
\label{sec:regulation}
\label{subsec:adapt-overhead}
\label{sec:overhead}

\para{Dynamic SM allocation}
\label{sec:narrowing}
\agreed{\sys is built with the recognition that common high-level concurrency tuning knobs, such as through the CPU core and GPU SM allocations, do not effectively protect CPU PTs, as explained in Section~\ref{subsec:contention}. 
However, within the framework of \sys's model-based pricing and fine-grained GPU job admission, we re-examined the potential of dynamic SM allocation, through the relatively recent NVIDIA green context mechanism.}

\agreed{We observe that the effect of reducing the SM footprint \agreed{(referred to as \emph{SM narrowing} in our discussion)} depends on the GPU workload.
Intuitively, it slows compute-heavy operators substantially. 
A bandwidth-bound decode job, on the other hand, can retain much of its speed when narrowed to fewer SMs. 
Interestingly, it \dkg{can maintain} most of its throughput while \dkg{reducing} CPU delay per GB of its GPU traffic, in other words, a flatter interference slope $\kappa_i$.
This is because that fewer SMs would reduce the intensity of concurrent GPU memory accesses, allowing more breathing room for CPU tasks. 
\dkg{A lower interference slope $\kappa_i$ can increase the allowance and let the budget $B_{i,j}(t)$ accrue faster, allowing earlier GPU admission.}}

\sys{} therefore runs prefill at full width and retains full-width CUDA graphs for the VLM, including its compute-heavy vision-encoder \atoms.
\agreed{For continuous LLM decode jobs \dkg{overlapping active CPU jobs}, %
\sys{} \dkg{uses} a fixed SM width (typically 8 or 16), based on its quick profiling during warm-up. \dkg{CPU-free decode uses all SMs.}
Figure~\dkg{\ref*{fig:breakdown}(a)} in the next Section gives related evaluation results.}

\agreed{
This allows \sys to expand its solution to a 2-D space of admission/scheduling and SM footprint management.
While a more comprehensive investigation is planned for our future work, \dkg{these} initial results reveal that under careful task decomposition and admission, mechanisms such as green context could deliver \dkg{a smoothing effect on GPU bandwidth consumption under edge CPU--GPU co-running}, not previously observed at inference engines' native scheduling granularity.
}

%
%
%

\para{Grant pipelining}
To hide the scheduler round trip, \sys{} authorizes and delivers one successor grant while the current grant runs.
At the next grant boundary, the engine checks the cached grant's work identity, lease and CPU-release limit, and discards it if stale.
A successor delivered ahead for a CPU release waits until that release before it can start.
%

%

%
%
%
%
%
%
%
%
%
%
%
%
%
%
%
%
%
%
%
%
%
%
\begin{figure*}[!t]
\centering
\includegraphics[width=\textwidth]{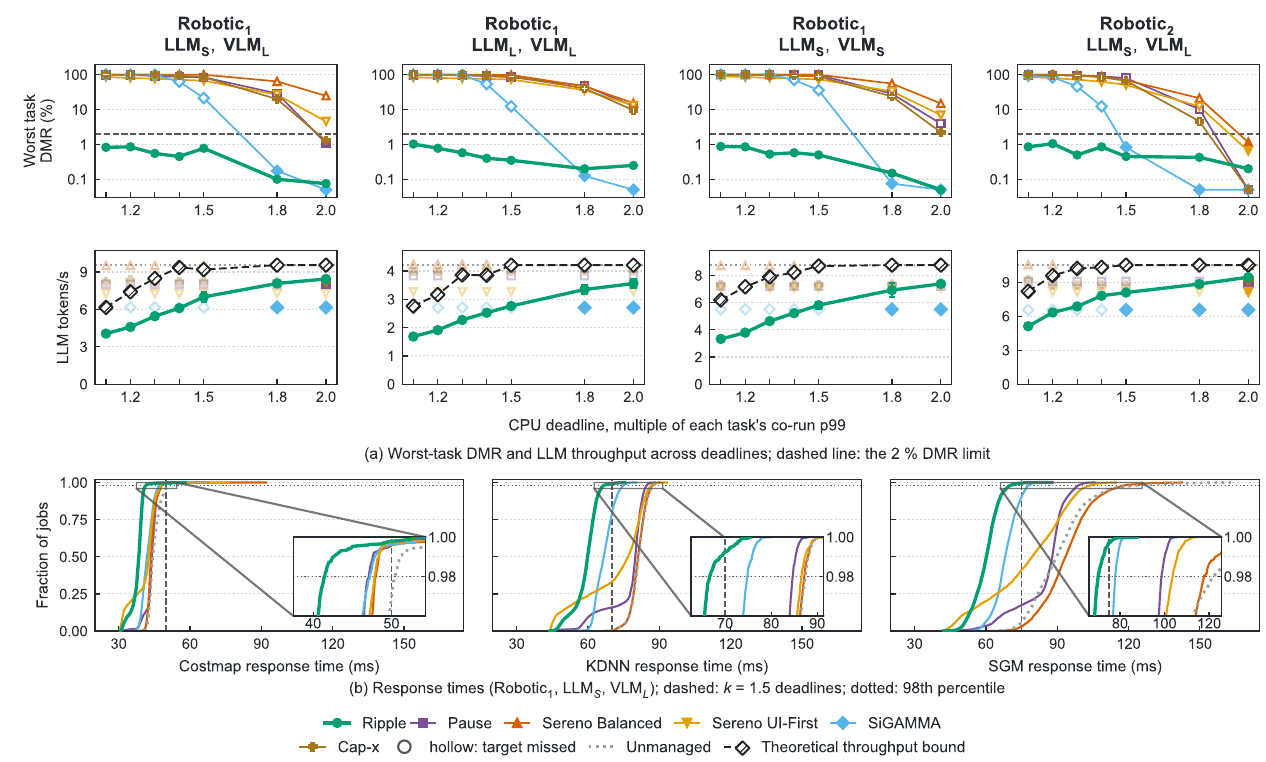}
\caption{\textbf{End-to-end results.} %
(a)~Worst-task DMR of CPU PTs %
and LLM tokens/s;
hollow markers indicate a CPU DMR limit or VLM deadline was missed. 
(b)~Response times with \jd{deadlines at $k=1.5$} on the \workloadmix{1}{S}{L} mix; each inset
magnifies where the curves cross the 98th percentile. 
}
\label{fig:eval}
\label{fig:cdf}
\end{figure*}

\begingroup
\color{black}
\renewcommand{\pagebudget}[1]{}
\renewcommand{\dkg}[1]{#1}
\section{Evaluation\pagebudget{2.75}}
\label{sec:evaluation}
We implemented \sys's three modules (engine hooks, scheduler, and estimator) in Python, with approximately 10.5k, 3.7k, and 2.2k LoC, respectively.

Our evaluation examines four properties: (1) \sys's capability to meet the CPU and VLM targets across representative CPU and GPU workload combinations (Section~\ref{sec:eval-feasible}); (2) its capability to retain GPU-side inference throughput while meeting given DMR limits (Section~\ref{sec:eval-ablation}); (3) its overhead and adaptation to changes (Section~\ref{sec:eval-ablation}); and (4) the impact of individual design/optimization (Section~\ref{sec:eval-ablation}).

\subsection{Experimental Setup}
\label{sec:evaluation-contracts}
\para{Testbed} We evaluate on an ASUS Ascent GX10, an NVIDIA DGX Spark system. Its GB10 SoC has a 20-core Arm CPU (10 Cortex-X925 and 10 Cortex-A725 cores) and a Blackwell GPU with 48 SMs, which share 128\,GB of LPDDR5X at 273\,GB/s. The machine runs Ubuntu 24.04.3 LTS with Linux 6.14, NVIDIA driver 580.95.05 and CUDA 13.0, and SGLang 0.5.12.

\para{Workloads and configurations}
\agreed{
Our evaluation runs three robotic perception/control tasks on the CPU (from those listed in Table~\ref{tab:workloads}), with an LLM and a VLM inference engine on the GPU. 
We create two CPU workload mixes, $Robotic_1$ (Costmap, KDNN, and SGM), and $Robotic_2$ (Control, ORB, and SGM). 
For LLM and VLM, we test two model configurations each:
$LLM_L$ with Qwen3-32B-FP8, 
$LLM_S$ with Qwen3-8B, 
$VLM_L$ with Qwen3-VL-30B-A3B-FP8, and
$VLM_S$ with Qwen3-VL-8B~\cite{qwen3,qwen3vl}.
\agreed{We evaluate four workload mixes to examine how workload composition affects protection and throughput.}
}

\dkg{CPU PTs use 100\,ms periods, following KITTI's 10\,Hz sensor rate~\cite{kitti}, except Control, which runs every 20\,ms, a 50\,Hz cadence used in embedded quadrotor MPC~\cite{salzmann2023neuralmpc}.
We set the deadlines of Costmap, KDNN, and SGM to $k$ times their co-run p99, measured with the other CPU tasks running and both GPU engines resident but idle, and sweep $k$ from 1.1 to 2.0 to cover different amounts of timing slack.
For detailed experiments, we select $k=1.5$, giving these tasks 50\% slack over their co-run p99; results at other settings are labeled separately.
Control and ORB use deadlines equal to their periods, requiring completion within the control cycle or before the next frame; the $Robotic_2$ sweep therefore varies only SGM's deadline.
In our experiments, the VLM period and deadline are both 5\,s for $VLM_L$ and 10\,s for $VLM_S$; the LLM runs continuously without a deadline.}

We use a DMR limit of 2\% for each CPU task in the scored window and require every VLM frame to meet its deadline.
We report mean throughput with one-standard-deviation error bars (often invisible).
\para{Baselines}
As the first framework designed for real-time CPU tasks that share a unified memory with GPU inference, we have a hard time identifying baseline systems.
Though there are no direct competitors, we do a best-effort comparison with several existing systems or alternative strategies:
(1) \emph{Sereno}~\cite{sereno}, designed for mobile SoCs that run background model inference beside foreground apps, pauses the LLM between decode steps to yield time shares to UI tasks (we run its ``Balanced'' and ``UI-First'' policies);
(2) \emph{SiGAMMA}~\cite{sigamma} uses a high-priority blocking kernel to suppress GPU memory traffic while a PT job is active;
(3) \emph{Pause} pauses both engines while any PT job is active, using SGLang's own pause command, which takes effect after the forward pass in flight;
(4) \agreed{\emph{Cap-$x$}} uses the same command and gives each PT job a \agreed{budget fraction $x$ of its solo median time} to tolerate the engines' co-run ($x=1/8, 1/4, 1/2$), pausing the latter whenever a active PT job has spent its budget;

\emph{Sereno} and \emph{SiGAMMA} run using their published
configuration, untuned, on SGLang 0.5.12 with the same engine flags used by \sys.
\jd{Only \sys{} react to the deadline.}
For reference, we also have the \emph{unmanaged} strategy, where all GPU-side engine jobs are admitted at once while the PT workloads run unprotected.

\endgroup

\subsection{Effectiveness in CPU PT Protection}
\label{sec:eval-feasible}
\label{sec:eval-mechanism}
\agreed{Figure~\ref{fig:eval} (a) gives \sys's overall performance, with 4 CPU-GPU workload mixes, at different deadline scaling factors ($k$ along the $x$ axis). 
Markers are filled only when \emph{all CPU tasks} meet the 2\% DMR limit and every VLM frame meets its deadline, otherwise hollow.}

\agreed{
The first row plots the worst deadline miss rate (DMR) across all three CPU PTs ($y$ axis in log scale). 
\sys is the \emph{only} solution meeting the DMR requirement consistently. 
A distant second is SiGAMMA, getting below the 2\% target with the most relaxed 2-3 deadline settings, due to its \agreed{partial suppression of inference through time-slicing with its blocking kernel}.
All the other baselines not only miss the target in all or nearly all tests, but produce close-to-100\% DMRs, \agreed{with the worst CPU task missing most of its deadlines even at $k=1.5$}.
This indicates existing approaches' failure in regulating the GPU-side aggressive memory bandwidth usage. 
In particular, the Sereno strategies fail to \agreed{respond to the remaining slack of individual CPU jobs, since their controller observes the LLM's slowdown rather than CPU deadlines}.
Even the pause strategy, which is in theory the harshest on the GPU workloads, is unable to contain the inference engines' memory traffic due to \agreed{the pause request taking effect only after the in-flight forward pass completes}.
}

\agreed{
The second row plots the GPU-side throughput, by LLM token generation speed. 
Here we add a theoretical throughput bound (diamonds), computed by replacing each GPU work \atom with a fixed-rate memory-access stream with the same volume. 
Therefore it also removes \sys overheads associated with discrete grants, VLM deadlines, and the scheduling protocol.
As expected, \sys's result trails behind this theoretical throughput bound, due to the conservative nature of its budget-based admission, its safety margins, as well as \sys's overhead.
Meanwhile, it clearly out performs any solutions that meet the DMR requirement (SiGAMMA, which delivers lower LLM throughput due to its 48-SM ``blocking kernel'' wasting GPU resources even with relatively safe CPU deadlines).
With more relaxed deadlines ($k$=1.8 or higher), it delivers higher LLM throughput than a few approaches that yields much higher DMR, 
as it admits more GPU work as deadline slack increases, whereas these baselines retain deadline-independent throttling}.

Among the four CPU-GPU workload combinations, the $LLM_L + VLM_L$ GPU combination is most aggressive in bandwidth use, rendering most baselines miss the DMR target significantly, even with the most forgiving 2$\times$ deadline setting.
$Robotic_2$, meanwhile, is slightly less vulnerable than $Robotic_1$, due to its lower bandwidth demand (Control and ORB has a per-job data volume of 10s of MBs, while Costmap and KDNN are at \jd{100MB-GB} level.)
Overall, the performance behavior of \sys and the baselines are quite consistent across the board.

\agreed{
Figure~\ref{fig:cdf}(b) shows PT response-time CDFs at $k=1.5$ for $Robotic_1$, $LLM_S$, and $VLM_L$; the vertical dashed lines mark deadlines.
Under \sys{}, the 98th-percentile response times of Costmap, KDNN, and SGM fall 8.6, 3.4, and 4.9\,ms below their respective deadlines, significantly outperforming the baselines.
The shapes of the CDF curves here also clearly reflect the vulnerability of each CPU PT, with SGM most sensitive to GPU bandwidth occupation.
}

\begin{figure}[!htb]
\centering
\includegraphics[width=\columnwidth]{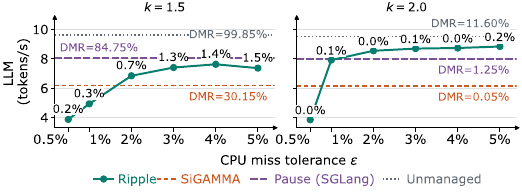}
\caption{Sensitivity to $\varepsilon$ on the \workloadmix{1}{S}{L} mix. Labels show worst-task CPU DMR.}
\label{fig:eval-epsilon}
\end{figure}

\agreed{Figure~\ref{fig:eval-epsilon} examines \sys's capability to work under varied DMR targets, with $\varepsilon$ varied between 0.5 and 5\%, under $1.5\times$ and $2\times$ deadlines, respectively.
For reference, we provide the LLM throughput of three baselines, SiGAMMA, Pause (SGLang), and Unmanaged, who deliver flat lines as they are oblivious to deadlines.
\sys, on the other hand, is able to relax its GPU work admission and produces higher inference performance while staying within the DMR tolerance. 
Consistent with Figure~\ref{fig:eval}(a), \sys allows better LLM throughput in general, compared to baselines that meet the DMR target.
The dip of \sys throughput from 4\% to 5\% DMR target with $1.5\times$ deadline shows its sensitivity to misses, which cost bandwidth utilization with the admission decision adjusted more conservative.}

\begingroup
\setlength{\intextsep}{5pt}
\makeatletter
\patchcmd{\@makecaption}{\vskip 10pt}{\vskip 4pt}{}{}
\makeatother
\begin{figure}[H]
\centering
\includegraphics[width=.94\columnwidth]{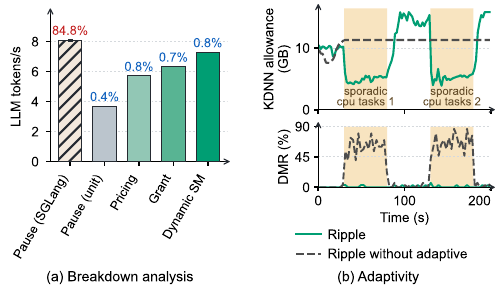}
\caption{Breakdown and adaptivity on the \workloadmix{1}{S}{L} mix \
}
\label{fig:evalrow}
\label{fig:breakdown}
\label{fig:adapt}
\end{figure}
\endgroup

\subsection{Breakdown Analysis}
\label{sec:eval-ablation}
\label{sec:eval-naive}
\label{sec:eval-estimator}
\para{Impact of individual techniques}
Figure~\ref{fig:breakdown}(a) gives the DMR and LLM throughput results by incrementally adding our proposed techniques: ``pricing'' (base-variation with pricing only), ``grant'' (issuing contiguous \atoms), and ``dynamic SM'' (SM narrowing with green context on LLM decode jobs). 
For comparison, we have two versions of the ``Pause'' strategy, one native from SGLang (pausing entire forward passes), the other at our refined \atom level. 

The SGLang native Pause significantly exceeds the 2\% DMR limit (84.8\%), although it stops the GPU whenever a CPU job runs.
When pausing at \atom level, it becomes extra-conservative.
With each of its optimizations, the green \sys bars safely stay under the 2\% DMR limit to protect the CPU PTs, while allowing more efficient bandwidth utilization by the GPU tasks, finally reaching 90.4\% of the SGLang Pause's throughput.

\para{Adaptivity}
With its pricing model observing the CPU PTs' performance, \sys is able to adapt to variability due to CPU-side contention or thermal throttling. 
The latter factor is beyond our control while the former can be tested. 
Figure~\ref{fig:adapt}(b) evaluates \sys by injecting two sporadic CPU tasks (memory readers) during a 200s run.  
The top plot shows how the KDNN's allowance automatically lowers during the two perturbation intervals and recovers thereafter, thanks to the $\bar a_{i,j}$ term in our pricing model. 
The bottom plot shows the DMR readings (binned at 3s), comparing \sys with a variation that has its adaptation component disabled (dashed line).
The difference is dramatic: KDNN's DMR averages at 0.6\% and 1.0\% in the two intervals, respectively, while the non-adaptive version reaches 60.4\% and 67.0\%.

\begin{figure}[!htb]
\centering
\includegraphics[width=.8\columnwidth]{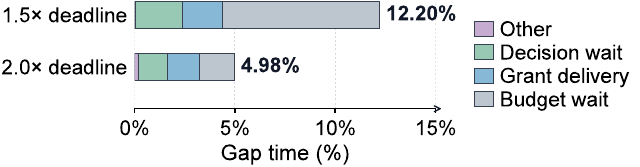}
\caption{Breakdown of inter-grant gap time as percentages of scored time on the \workloadmix{1}{S}{L} mix.
}
\label{fig:eval-overhead}
\end{figure}

\para{Overhead}
We then examine the management and scheduling overhead of \sys, by measuring the percentage of time during the GPU workloads' active execution that \sys has no grant running.   
Our detailed profiling reveals the breakdown of such ``gap time'', plotted in Figure~\ref{fig:eval-overhead}. 
With a more stringent deadline ($1.5\times$), the gap is dominated by ``budget wait'', the time \sys waits for sufficient budget to release the next \atom, a price it has to pay to regulate GPU bandwidth usage and protect CPU PTs. 
This item becomes much smaller with a $2\times$ deadline, as \sys schedules \atoms more aggressively. 
The other categories, such as \sys decision wait and delivery times, are its overhead. 
They also shrink when the deadline is relaxed, as there is a higher chance more \atoms are issued within each grant, reducing this cost. 
Together, such gap times contribute to the inference throughput loss against the ``unmanaged'' reference in Figure~\ref{fig:eval}. 
At $2\times$ deadline, only 5\% of GPU time is not running inference. 
\para{Other results} \jd{After a 3\,s GPU copy, allowances recovered within 20\,s for KDNN and SGM and 24\,s for Costmap in both runs.
Without a CPU-task profile, \sys{} misses no more deadlines in its first 10\,s than when started with one.
The same configuration meets the CPU DMR limits and VLM deadlines across all four workload mixes without per-workload tuning.
In two one-hour runs at $k=1.5$, every 100\,s window met CPU DMR limits, with no VLM deadline misses.}
\section{Related Work\pagebudget{0.5}}
\label{sec:related}
\outlinepoint{Real-time memory regulation}
Memory regulators use budgets or coordinated phases.
MemGuard~\cite{memguard} budgets per-core memory accesses, and RT-Gang~\cite{rtgang} throttles best-effort cores while a real-time gang runs.
BWLOCK++~\cite{bwlockpp} throttles CPU traffic to protect GPU kernels; SiGAMMA~\cite{sigamma} blocks GPU accesses during PREM~\cite{prem} CPU memory phases.
These require counters and throttling or phase separation.
\sys{} learns from CPU completions, admitting overlapping inference within per-job allowances without bandwidth counters or CPU memory-phase annotations.

\outlinepoint{Inference preemption and co-serving}
ConServe~\cite{conserve} preempts offline inference at layer boundaries, while Valve~\cite{valve} bounds preemption latency and rate to protect GPU requests.
Sereno~\cite{sereno} yields background NPU inference according to draft-subgraph latency to protect foreground applications.
\sys{} admits work at fine boundaries against each independent CPU job's declared deadline; inference latency alone does not reveal that job's slack.

\outlinepoint{GPU spatial partitioning}
libsmctrl~\cite{libsmctrl} and nvtaskset~\cite{nvtaskset} provide compute partitioning; CUDA green contexts~\cite{nvidiagreenctx} restrict a stream's kernels to selected SMs.
DuetServe~\cite{duetserve} adaptively divides SMs between prefill and decode to meet GPU latency targets.
\sys{} combines per-grant SM control through green contexts with temporal admission to protect CPU deadlines.
Narrowing can lower CPU interference per GPU byte and increase the allowance; throughput improves only when enough additional admitted work offsets slower execution.

\outlinepoint{Prediction and feedback}
Clockwork~\cite{clockwork} predicts inference duration, SOLA~\cite{sola} updates latency estimates using measured-to-predicted ratios, and FC-EDF~\cite{fcedf} adjusts service levels and admission from deadline-miss feedback.
\sys{} combines duration prediction with a model of CPU delay per GPU byte and adaptive conformal inference~\cite{gibbs2021aci} for its CPU prediction margin (Appendix~\ref{app:pricing}); together they turn deadline slack into a GPU-byte allowance.

\section{Conclusion and Future Work\pagebudget{0.25}}
\label{sec:conclusion}
\cb{\sys demonstrates that with UMA edge AI devices, the CPU can accommodate diverse time-critical robotic/agentic workloads while contemporary LLM/VLM inference runs on the GPU.
We protect the less intensive, but highly sensitive CPU-side memory accesses by leveraging semantic knowledge as well as available resource management mechanisms on the GPU workloads to overcome the lack of bandwidth counters on these edge devices with unified memory.}

Two questions merit further study. First, narrowing the SM footprint can
reduce CPU interference per GPU byte, but the throughput benefit depends on the
workload and deadline slack. Jointly choosing the SM count and admission timing
could exploit this trade-off more fully. Second, long prefill operations call
for finer execution boundaries; the challenge is to shorten each uninterrupted
interval without excessive weight rereads or scheduling overhead.

\label{body:end}

\bibliographystyle{mlsys2025}
\begingroup
\raggedright
\bibliography{refs}
\endgroup

\begin{samepage}
\section*{AI Usage Disclosure}
We used OpenAI Codex and Anthropic Claude iteratively under author direction to explore design
alternatives, discover and summarize literature, propose experimental contrasts,
implement and debug runtime and experiment-harness code, develop data-reduction
and figure-generation scripts, and draft, revise, and format manuscript text.
This assistance included executing author-authorized commands and helping
interpret retained experiment outputs. Reported hardware measurements are drawn
from experiment records rather than AI-generated numerical observations.
The authors retain responsibility for the design decisions, experimental
methodology, references, analysis, and all content of the paper.

\end{samepage}

\clearpage
\appendix
\makeatletter
\setlength{\@fptop}{0pt}
\setlength{\@fpsep}{12pt}
\setlength{\@fpbot}{0pt plus 1fil}
\makeatother
\renewcommand{\topfraction}{0.9}
\renewcommand{\textfraction}{0.1}
\renewcommand{\floatpagefraction}{0.9}

\section{Hardware and Interference Measurements}
\label{app:hardware-investigation}
\label{app:channel}

\para{Shared-memory contention.}
The selected co-run measurements give about
257\,GB/s for sequential CPU reads beside Qwen3-8B decode and
190\,GB/s for read--write copying, against the nominal 273\,GB/s.
Table~\ref{tab:bandwidth-accounting} summarizes how the rates are computed.
GPU decode traffic is calculated from
its output rate and per-token byte count; CPU traffic is calculated by the
benchmark from its processed data and elapsed time. In separate measurements,
Nsight Compute L2 read misses agree with checkpoint-based decode-byte counts
within 0.7\,\% for Qwen3-8B and Qwen3-32B in BF16 and AWQ. Beside Qwen3-8B decode,
a CPU thread that reads 30\,GB/s alone obtains 6.6\,GB/s, while the combined
rate remains about 50\,GB/s below the peak. Thus the peak is not a fixed
pool either processor can fully reclaim when the other withdraws.
In a separate experiment, an eight-thread CPU reader shares memory with a
GPU reader reduced by duty cycling, SM count, or arithmetic intensity.
The comparison uses three matched GPU rates.
The CPU recovers 0.401, 0.572, and 0.570\,GB/s per GB/s withdrawn,
respectively.

\begin{table}[tbp]
\caption{Inputs and calculations for the shared-memory bandwidth measurements.}
\label{tab:bandwidth-accounting}
\par\vspace{0.1in}
\centering\small
\begin{tabular}{@{}>{\raggedright\arraybackslash}p{0.29\columnwidth}>{\raggedright\arraybackslash}p{0.65\columnwidth}@{}}
\toprule
Quantity & Definition \\
\midrule
16.4 GB/token & Qwen3-8B weight bytes per batch-1 decode token used in the co-run calculation \\
GPU rate & Decode tokens/s $\times$ 16.4 GB/token \\
CPU rate & Benchmark bytes processed divided by elapsed time \\
Combined rate & Sum of GPU and CPU rates \\
273 GB/s & Nominal GB10 memory bandwidth \\
\bottomrule
\end{tabular}
\end{table}

\para{DRAM rather than cache interference.}
A CPU reader sustains 150--151\,GB/s beside GPU data fitting in the GPU's
24\,MB L2, but only 85--86\,GB/s when the GPU data spill to DRAM. Thrashing
the 16\,MB system-level cache does not shift this threshold. A dependent
CPU load chain's mean access latency rises from 126.1 to 283.7\,ns under a
GPU stream. Its last-level read miss ratio changes by only $+0.5\%$, and
memory accesses per instruction remain unchanged. A dependent integer
chain without memory traffic retains its rate. The dependent-load
experiment thus identifies longer memory service time rather than
additional cache misses or a slower CPU.

\para{Available interfaces.}
Table~\ref{tab:interface-audit} describes the tested firmware and software. The administrator-level inventory
finds CPU-core, statistical-profiling, and SMMU counters, but no uncore
or memory-controller PMU: firmware provides neither an APMT table for
those units nor an MPAM table, although the kernel has both drivers.
\begin{table}[tbp]
\caption{Memory and GPU control interfaces on the evaluated GB10 stack.}
\par\vspace{0.1in}
\centering
\small
\setlength{\tabcolsep}{4pt}
\begin{tabular}{@{}>{\raggedright\arraybackslash}p{0.32\columnwidth}>{\raggedright\arraybackslash}p{0.24\columnwidth}>{\raggedright\arraybackslash}p{0.37\columnwidth}@{}}
\toprule
Means & Effect on CPU & Shortcoming \\
\midrule
Uncore / memory-controller PMU & -- & Not exposed by the firmware \\
DCGM and Nsight DRAM metrics & -- & None for this GPU \\
CPU core PMU, GPU L2 counters & -- & One side's cache traffic only \\
MPAM, MIG, OS regulation & -- & Not enabled, or blind to the GPU \\
GPU clock 2.4$\to$1.1\,GHz & p50 $-7\%$ & Too small; locking halves decode \\
CPU clock & -- & Limit accepted, not applied \\
SMs 48$\to$24 & p50 $-9\%$, tokens $-14\%$ & Static, deadline-blind \\
CPU cores, engine priorities & CPU side / GPU queue only & Cannot move the GPU's share \\
\bottomrule
\end{tabular}

\label{tab:existing}
\label{tab:interface-audit}
\label{tab:knobs}
\end{table}

\para{CPU traffic and narrowing.}
\label{app:cpu-bytes}
With the GPU idle, core bus counters give about 1.1\,GB per Costmap job,
370\,MB for SGM, 250\,MB for KDNN, 11\,MB for ORB, and 2\,MB for Control.
The two smallest include cache hits and are upper bounds.
With Costmap, KDNN, and SGM together, average
CPU traffic is 18.7\,GB/s and the 1\,ms p99 is 64--65\,GB/s; their p99
completion times rise 2--7\% over running alone. In the synthetic
GPU-stream experiment, narrowing FP8 work to 16 of 48 SMs reduces its byte
rate by 8\% and CPU delay per GB by 9--35\%. BF16 still saturates memory
at 16 SMs, with harm per byte unchanged within run-to-run spread.

\section{GPU Volume Accounting and Execution}
\label{app:bytes}
\label{app:granularity}

\para{Operation granularity.} Table~\ref{tab:granularity} compares the
duration of complete forwards with individual operations: Qwen3-8B decode
takes 69--74\,ms per forward, while an operation spans 1.68--8.40\,ms.
Table~\ref{tab:implementation} identifies the operation boundaries used by
the two engines: \sys{} can admit one or more operations in a grant, with eager
execution for the LLM and breakable CUDA graphs for VLM decode.

\begin{table}[tbp]
\caption{Diagnostic CUDA-event spans on GB10, including host enqueue gaps.}
\par\vspace{0.1in}
\label{tab:granularity}
\centering
\footnotesize
\setlength{\tabcolsep}{3pt}
\renewcommand{\arraystretch}{1.05}
\begin{tabular}{@{}>{\raggedright\arraybackslash}p{0.36\linewidth}
  >{\raggedleft\arraybackslash}p{0.29\linewidth}
  >{\raggedleft\arraybackslash}p{0.29\linewidth}@{}}
\toprule
Model / phase & Whole forward\newline (ms) & Single operation\newline (ms) \\
\midrule
Qwen3-8B\newline Prefill & 76.61 & 1.83--7.37 \\
Qwen3-8B\newline Decode & 69.00--74.04 & 1.68--8.40 \\
Qwen3-VL\newline Image extend & 285.23 & 0.05--4.72 \\
Qwen3-VL\newline Decode & 19.48--21.10 & 0.02--4.69 \\
\bottomrule
\end{tabular}
\par\vspace{2pt}
\begin{minipage}{\linewidth}\small
Ranges show observed minima and maxima.
\end{minipage}
\end{table}

\begin{table*}[t]
\caption{Selected single-request SGLang paths for Qwen3-8B and
Qwen3-VL-30B-A3B-Instruct-FP8. Counts are registered operations per forward
pass; one grant can admit several.}
\par\vspace{0.1in}
\label{tab:implementation}
\centering
\small
\setlength{\tabcolsep}{4pt}
\renewcommand{\arraystretch}{1.10}
\begin{tabular}{@{}>{\raggedright\arraybackslash}p{0.22\textwidth}
  >{\raggedleft\arraybackslash}p{0.09\textwidth}
  >{\raggedright\arraybackslash}p{0.45\textwidth}
  >{\raggedright\arraybackslash}p{0.17\textwidth}@{}}
\toprule
Engine and phase & Operations & Boundary groups & Execution mode \\
\midrule
\mbox{Qwen3-8B Prefill} & 38 &
Input preparation; 36 decoder layers; output head & Eager \\
\addlinespace[2pt]
\mbox{Qwen3-8B Decode} & 37 &
36 decoder layers (preparation included with the first); output head & Eager \\
\addlinespace[2pt]
\mbox{Qwen3-VL Image/prompt} & 108 &
Preparation; vision input, attention/MLP regions, and merger; language input,
48 layers, norm, and output & Eager \\
\addlinespace[2pt]
\mbox{Qwen3-VL Decode} & 52 &
Preparation; language input, 48 layers, norm, and output &
Breakable CUDA graphs \\
\bottomrule
\end{tabular}
\end{table*}

\begin{figure}[!tbp]
\centering
\includegraphics[width=\columnwidth]{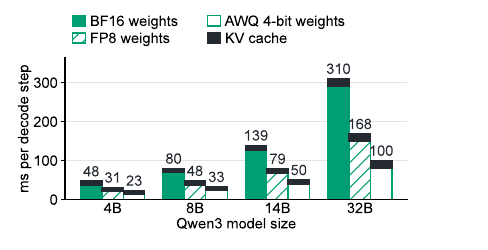}
\caption{Time one batch-1 decode step takes at a 16{,}384-token context when its bytes
stream at 220\,GB/s. Appendix~\ref{app:bytes} gives the method.}
\label{fig:bytes}
\end{figure}

\para{Byte charges.}
A decode step's weight and KV traffic grows with model size and context length;
Figure~\ref{fig:bytes} converts the counted bytes at a 16,384-token context
to step time using a 220\,GB/s streaming rate.
A decoder layer at sequence length $T$ is charged
$W_\ell+A_\ell+(T+1)K$: weights, activation traffic, and KV reads for $T$
positions plus one write. Counts come from checkpoint tensors and the
engine's sequence length; the output head, final norm, and embedding row
are charged separately. Qwen3-8B has $W_\ell=0.386$\,GB and $K=4096$ bytes,
so each thousand context tokens adds about 1\% to step bytes. A short-context
step reads 15.19\,GB in 69--74\,ms, or 205--220\,GB/s. MoE decode charges
the router and selected experts; image prefill charges all experts.

One-metric-per-run Nsight Compute checks match computed decode bytes to L2
read misses within 0.7\% for Qwen3-8B/32B in BF16 and AWQ; a known-size
stream agrees within 0.02\%. An initial VLM check corrected its per-step
charge by 17\%. Subsequent Qwen3-32B-FP8 and Qwen3-VL-30B-A3B/8B checks
agree within 0.5\% for decode, 0.4\% for routed/all-expert MoE charges, and
1\% for a frame. These checks are offline; admission reads no counter.
Prefill can reread weights across GEMM tiles: Qwen3-8B reads 1.6 times its
weights at 1024 input tokens and 18.9 times at 16,384, versus 1.009 for
the evaluation's 37-token prompt.

\begin{figure}[!t]
\centering
\IfFileExists{figures/06-cuda-graph-ablation.pdf}{%
  \includegraphics[width=\columnwidth]{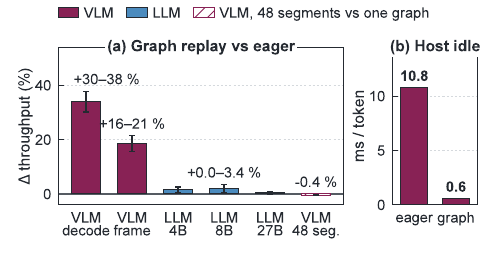}%
}{%
  \fbox{\parbox[c][\eightlinefigureheight][c]{0.93\columnwidth}{\centering\small
  \textbf{Figure pending:} \texttt{figures/06-cuda-graph-ablation.svg}\\[2pt]
  Layout placeholder, no data drawn.}}%
}
\caption{CUDA-graph ablation, engine alone on GB10. (a)~Throughput change of graph replay
over eager execution, one contrast per bar; the hatched bar is 48 graph segments
per decode step against one. (b)~Host-dispatch idle per VLM decode token.}
\label{fig:cuda-graph-ablation}
\end{figure}

\para{Execution boundaries.}
\phantomsection\label{app:execution-path}
LLM execution is eager; VLM decode uses SGLang's breakable CUDA graphs,
with eager image processing and prefill. Figure~\ref{fig:cuda-graph-ablation}
compares graph replay with eager execution in each engine alone. Replay speeds
VLM decode by 30--38\% and a frame by 16--21\% in engine-only tests,
versus 0--3.4\% for the tested dense LLMs. Segmenting a graph without
admission callbacks increases a fixed frame sequence's duration by 0.4\%.
VLM decode issues roughly 1000 kernels averaging 28\,$\mu$s; tested LLM
kernels average 166\,$\mu$s. A CUDA event marks grant completion, so reports
measure execution rather than submission. Stream dependencies preserve
serial order when switching SM widths. Separate engine processes time-slice
on the tested GB10 stack; restricting one does not reserve SMs for the other.
In a single-process BF16 matrix-kernel microbenchmark, concurrent decode-like
streams took 0.9--1.6\% longer than serial execution;
a decode-like stream paired with a prefill-like stream took 5.4--5.7\% longer.

\section{Pricing and Estimator Details}
\label{app:pricing}
\label{app:estimator}

Each task's model uses recursive least squares with exponential forgetting:
effective windows of $W=50$ jobs for the intercept and $W_\kappa=200$ for the
slope. The held slope changes only when the new estimate is positive and at
least two standard errors above zero. The adaptive margin uses the last $W_q=200$ residuals,
starts after 50 samples, and uses one adaptive level across VLM phases.
Before then, solo profiles seed the fit and margin; the warm-up margin uses
the larger of the phase and pooled quantiles over up to 50 residuals. The
phase split separates releases in the first 10\,\% of a VLM period from the rest.

The conservative CPU duration is $\bar a_{i,j}=\hat a_{i,j}+q_{i,j}$:
the fitted interference-free duration plus a residual margin.

\para{Margin.} The margin is a rank among the $n_{i,j}\le W_q$ retained residuals whose level $\alpha_{i,j}$ adapts after every job,
\begin{equation}
\begin{aligned}
q_{i,j} &= \hat e_{(m)}, \quad m = \bigl\lceil (1-\alpha_{i,j})(n_{i,j}+1) \bigr\rceil,\\
\alpha_{i,j+1} &= \alpha_{i,j} + \eta\,(\varepsilon'_i - \mathrm{err}_{i,j}),
\end{aligned}
\label{eq:margin}
\end{equation}
with $\eta=0.01$, where $\mathrm{err}_{i,j}=1$ if the job exceeds its saved release-time pricing line $\hat a_{i,j}+q_{i,j}+s_{i,j}C_{i,j}$; an unfinished job exceeds any finite line. Each residual $\hat e_{i,j}=CT_{i,j}-\hat a_{i,j}-\hat\kappa_{i,j}C_{i,j}$ is computed before its job enters the fit.
This adaptive conformal inference update~\cite{gibbs2021aci,angelopoulos2023gentle} widens the margin by lowering the level when a job runs longer than its saved prediction plus margin, and otherwise raises it gradually. The feedback error and the deadline miss $CT_{i,j}>D_i$ are scored separately; rank endpoints are specified in Appendix~\ref{app:strict}.

\para{Window target.} To leave room for variation within the contract's window (Section~\ref{sec:overview}), \sys{} sets
$\varepsilon'_i = \max(10^{-6},\ \varepsilon_i - z_\varepsilon\sqrt{\varepsilon_i(1-\varepsilon_i)/n_i})$ with $z_\varepsilon = \Phi^{-1}(1-\varepsilon_i)$ and $n_i=\max(1,\lfloor L/T_i\rfloor)$ for $L=100$\,s. The subtraction uses a normal approximation to an independent binomial count~\cite{casella2002} as the reference buffer; actual window misses are evaluated from job outcomes.
With $\varepsilon_i = \delta = 2\,\%$, $\varepsilon'_i$ is 1.09\,\% for the 100\,ms tasks ($n_i = 1{,}000$) and 1.59\,\% for the 20\,ms Control task in $Robotic_2$ ($n_i = 5{,}000$).

\para{Pessimistic slope.} Pricing reserves slack for residual variation and slope uncertainty.
The margin covers the scatter of jobs around the fit of Eq.~\ref{eq:victim} at the GPU volumes they received, but not the error of the slope itself, which grows with the volume.
The uncertainty estimate is $\kappa^u_{i,j} = \max(0,\hat\kappa^{\mathrm{LS}}_{i,j}) + z_\varepsilon\,\mathrm{SE}_{i,j}$ with $z_\varepsilon = \Phi^{-1}(1-\varepsilon_i)$, using the current least-squares slope. Pricing uses $s_{i,j}=\max(\hat\kappa_{i,j},\kappa^u_{i,j})$, where $\hat\kappa_{i,j}$ is the held slope, or $s_{i,j}=\hat\kappa_{i,j}/0.85$ until a positive uncertainty estimate is available. Equivalently, $\beta_{i,j}=\hat\kappa_{i,j}/s_{i,j}$ is the share of residual slack used for GPU volume.
The cap $A^{\max}_i = \rho D_i$ uses the profiled peak rate $\rho$: the larger of the highest per-\atom byte rate and the rate measured with both engines decoding concurrently.

\para{Issuing the allowance.} The nominal allowance is rate-limited to a
base that changes by at most $\pm25\,\%$ per release, using one \atom as the
growth reference below that size, and remains capped at $A_i^{\max}$.
Downward rate limiting can retain bytes above the new nominal allowance.
While slope uncertainty exceeds 10\,\% or is not yet estimable, one job in
eight of the task with the smallest allowance receives 40\,\% fewer bytes.
The one-\atom probes below also affect the issued $A_{i,j}$. A best-effort
grant before a deadline-bound request's due time must finish 1\,ms before it.

\para{Pacing.} Releasing a job's allowance linearly over $\hat a_{i,j}$
(Section~\ref{sec:cpu-admission}) is what budget-pacing regulators do~\cite{aghilinasab2020}.
With floor $f$, the budget is
\begin{equation}
\begin{split}
B_{i,j}(t)={}&\min\!\left(A_{i,j},\max\!\left(f,
\frac{A_{i,j}(t-R_{i,j})}{\hat a_{i,j}}\right)\right)\\
&-C_{i,j}(t),
\end{split}
\end{equation}
with $\hat a_{i,j}$ floored at 1\,ns. An allowance below one \atom receives a one-\atom probe on alternate releases, subject to the guard below. The probe size is the same for every task: the largest, over the bound engines, of each engine's smallest nonzero decode \atom, so that every engine can issue one \atom of its steady-state work under it (385{,}978{,}880 bytes, one Qwen3-8B layer, on \workloadmix{1}{S}{L}); a job issued exactly
$A^{\max}_i$, because its raw allowance exceeds it, has its whole allowance released at
entry. Inherited grants remain charged separately, as described below.

\para{Charging.} Bytes are charged at admission and never refunded once a
grant has executed, so splitting work into smaller grants does not relax a
job's limit; a grant returned unexecuted is refunded. With no CPU job active,
no \budget limits GPU work.

\para{The crossing bound.} The crossing check of Section~\ref{sec:cpu-admission} models
the response time of a job $PT_{i,j}$ released at $R_{i,j}$ under a grant
with predicted end $t^{\mathrm{end}}$ and $c$ bytes as
$(t^{\mathrm{end}}-R_{i,j}) + \max(\hat a_{i,j} + q_{i,j}, p_i) + \hat\kappa_{i,j} \max(0, \bar A_i - c)$:
after the grant the job can receive at most $\bar A_i - c$ further bytes, and
$\max(\cdot, p_i)$, with $p_i$ the task's solo p99, keeps a low intercept
from shrinking its busy time. $\bar A_i$ bounds the allowance the job can be
issued whatever completes before $R_{i,j}$: it includes the 25\,\% growth rule
and the one-\atom probe, both capped at $A^{\max}_i$. A job's charge can still exceed
$A^{\max}_i$ by the grants in flight that it inherits at its release. Capping
the crossing at $T_i$ keeps a grant from reaching the task's next release.

\para{Rank endpoints and the tight-deadline guard.}
\label{app:strict}
If the rank exceeds the retained sample count, the margin update uses $q_{i,j}=D_i$ to represent an unbounded margin; a saved $q_{i,j}\ge D_i$ records
$\mathrm{err}_{i,j}=0$. At $\alpha_{i,j}\ge1$ it uses the smallest residual.
The updates use integer parts per million. Jobs assigned this margin follow the same
rate limit and probe rules, and their deadline outcomes are scored normally.
At the tightest deadlines \sys{} adds one
rule: if the predicted delay from GPU volume $f$ exceeds a job's remaining slack, that is,
$\lceil \hat\kappa_{i,j} f \rceil > D_i - \max(\hat a_{i,j} + q_{i,j}, p_i)$, evaluated
with the estimates that price it, then that job gets neither the probe nor the floor $f$
nor a crossing into its release.

\para{Task exemptions.} The model-based certification check requires
$\kappa^u_{i,j}\rho<1$ and
\[
\begin{aligned}
\frac{h_{i,j}+2\kappa^u_{i,j}\rho H}{1-\kappa^u_{i,j}\rho}
&\le \min\bigl(D_i,\ \hat a_{i,j}+q_{i,j}\\
&\qquad +\beta_{i,j}[D_i-\hat a_{i,j}-q_{i,j}]_+\bigr),
\end{aligned}
\]
where $h_{i,j}=\hat a_{i,j}+q_{i,j}+[\hat\kappa_{i,j}-\kappa^u_{i,j}]_+\bar C_i$
reanchors the intercept at its fit's weighted mean charge $\bar C_i$.
After 50 consecutive passes the task is exempt. A failed check or an observed
response above the saved line $h_{i,j}+\kappa^u_{i,j}C$ ends exemption; a
deadline miss while exempt, or a new lower slope estimate
$\hat\kappa^{\mathrm{LS}}_{i,j}-2\mathrm{SE}_{i,j}$ above the previous
$\kappa^u_{i,j}$, disables exemption for the rest of the run.

\section{Supplementary Evaluation}
\label{app:eval}
\label{app:protocol}

\para{Protocol and baselines.}
The \sys{} grid uses 140\,s preparation; the external baselines use 30\,s.
Both use a 100\,s score. A point
is feasible only if every run meets every CPU task's 2\% DMR limit
and every VLM deadline; all \sys{} grid runs meet these criteria.
Each workload mix covers seven deadline settings.
The controller uses the same parameter
settings across mixes, with the SM width chosen automatically during
warm-up. Table~\ref{tab:eval-full}
reports LLM throughput and the highest pooled CPU-task DMR for every deadline
and workload. For example, at $k=1.5$ on \workloadmix{1}{S}{L}, \sys{}
produces 7.00 tokens/s at 0.78\% DMR, while SiGAMMA produces 6.16 tokens/s
at 21.50\% DMR.

\begin{table*}[!t]
\caption{End-to-end results by workload and deadline. Each cell gives LLM throughput (tokens/s) / CPU DMR (\%). DMR is the highest task miss rate after pooling jobs across runs. Cap-$x$ selects the highest-throughput level meeting the CPU and VLM targets, or the lowest-DMR level if none meets them.}
\label{tab:eval-full}
\par\vspace{0.1in}
\centering\small
\setlength{\tabcolsep}{2pt}
\begin{tabular}{@{}ll*{7}{>{\centering\arraybackslash}p{1.70cm}}@{}}
\toprule
Mix & Scheduler & $1.1\times$ & $1.2\times$ & $1.3\times$ & $1.4\times$ & $1.5\times$ & $1.8\times$ & $2.0\times$ \\
\midrule
\multirow{7}{*}{\shortstack[l]{$Robotic_1$\\$LLM_S$\\$VLM_L$}} & Ripple & \mbox{4.06/0.82} & \mbox{4.59/0.85} & \mbox{5.45/0.55} & \mbox{6.10/0.45} & \mbox{7.00/0.78} & \mbox{8.08/0.10} & \mbox{8.44/0.08} \\
 & Sereno UI-First & \mbox{7.25/85.50} & \mbox{7.25/80.47} & \mbox{7.25/76.15} & \mbox{7.25/71.67} & \mbox{7.25/66.95} & \mbox{7.25/26.60} & \mbox{7.25/4.57} \\
 & Sereno Balanced & \mbox{9.53/100.00} & \mbox{9.53/100.00} & \mbox{9.53/100.00} & \mbox{9.53/100.00} & \mbox{9.53/99.85} & \mbox{9.53/63.95} & \mbox{9.53/24.78} \\
 & SiGAMMA & \mbox{6.16/99.98} & \mbox{6.16/99.80} & \mbox{6.16/98.83} & \mbox{6.16/62.92} & \mbox{6.16/21.50} & \mbox{6.16/0.18} & \mbox{6.16/0.02} \\
 & Pause & \mbox{8.03/98.92} & \mbox{8.03/98.65} & \mbox{8.03/91.25} & \mbox{8.03/86.33} & \mbox{8.03/84.20} & \mbox{8.03/28.93} & \mbox{8.03/1.08} \\
 & Cap-$x$ & \mbox{8.21/99.98} & \mbox{8.37/99.88} & \mbox{8.21/99.55} & \mbox{8.21/93.42} & \mbox{8.21/84.05} & \mbox{8.21/19.83} & \mbox{8.21/1.28} \\
 & Unmanaged & \mbox{9.55/100.00} & \mbox{9.55/100.00} & \mbox{9.55/100.00} & \mbox{9.55/100.00} & \mbox{9.55/99.72} & \mbox{9.55/48.77} & \mbox{9.55/17.27} \\
\midrule
\multirow{7}{*}{\shortstack[l]{$Robotic_1$\\$LLM_L$\\$VLM_L$}} & Ripple & \mbox{1.68/1.02} & \mbox{1.91/0.78} & \mbox{2.27/0.58} & \mbox{2.52/0.40} & \mbox{2.76/0.35} & \mbox{3.34/0.20} & \mbox{3.56/0.25} \\
 & Sereno UI-First & \mbox{3.27/83.50} & \mbox{3.27/80.75} & \mbox{3.27/78.28} & \mbox{3.27/75.75} & \mbox{3.27/72.50} & \mbox{3.27/36.02} & \mbox{3.27/13.15} \\
 & Sereno Balanced & \mbox{4.23/100.00} & \mbox{4.23/99.88} & \mbox{4.23/99.67} & \mbox{4.23/99.48} & \mbox{4.23/98.08} & \mbox{4.23/46.65} & \mbox{4.23/15.30} \\
 & SiGAMMA & \mbox{2.71/100.00} & \mbox{2.71/99.50} & \mbox{2.71/98.05} & \mbox{2.71/53.68} & \mbox{2.71/12.35} & \mbox{2.71/0.12} & \mbox{2.71/0.00} \\
 & Pause & \mbox{3.83/98.85} & \mbox{3.83/98.10} & \mbox{3.83/95.73} & \mbox{3.83/90.92} & \mbox{3.83/84.70} & \mbox{3.83/47.75} & \mbox{3.83/12.60} \\
 & Cap-$x$ & \mbox{3.98/99.78} & \mbox{3.98/99.62} & \mbox{3.98/99.30} & \mbox{4.01/96.30} & \mbox{3.98/89.10} & \mbox{4.01/40.30} & \mbox{4.01/9.50} \\
 & Unmanaged & \mbox{4.21/100.00} & \mbox{4.21/99.95} & \mbox{4.21/99.70} & \mbox{4.21/99.52} & \mbox{4.21/98.10} & \mbox{4.21/62.00} & \mbox{4.21/32.34} \\
\midrule
\multirow{7}{*}{\shortstack[l]{$Robotic_1$\\$LLM_S$\\$VLM_S$}} & Ripple & \mbox{3.33/0.88} & \mbox{3.79/0.85} & \mbox{4.64/0.52} & \mbox{5.22/0.58} & \mbox{5.81/0.50} & \mbox{6.92/0.15} & \mbox{7.38/0.02} \\
 & Sereno UI-First & \mbox{7.08/89.85} & \mbox{7.08/85.45} & \mbox{7.08/80.90} & \mbox{7.08/77.38} & \mbox{7.08/73.12} & \mbox{7.08/33.27} & \mbox{7.08/6.90} \\
 & Sereno Balanced & \mbox{8.72/100.00} & \mbox{8.72/100.00} & \mbox{8.72/100.00} & \mbox{8.72/100.00} & \mbox{8.72/99.78} & \mbox{8.72/55.45} & \mbox{8.72/14.92} \\
 & SiGAMMA & \mbox{5.50/99.98} & \mbox{5.50/99.75} & \mbox{5.50/99.12} & \mbox{5.50/70.55} & \mbox{5.50/35.92} & \mbox{5.50/0.08} & \mbox{5.50/0.02} \\
 & Pause & \mbox{7.21/99.90} & \mbox{7.21/99.90} & \mbox{7.21/99.80} & \mbox{7.21/99.70} & \mbox{7.21/98.65} & \mbox{7.21/28.30} & \mbox{7.21/3.93} \\
 & Cap-$x$ & \mbox{7.31/99.90} & \mbox{7.31/99.78} & \mbox{7.31/98.78} & \mbox{7.31/93.58} & \mbox{7.31/88.38} & \mbox{7.37/23.65} & \mbox{7.41/2.22} \\
 & Unmanaged & \mbox{8.76/100.00} & \mbox{8.76/100.00} & \mbox{8.76/100.00} & \mbox{8.76/100.00} & \mbox{8.76/99.72} & \mbox{8.76/50.40} & \mbox{8.76/18.55} \\
\midrule
\multirow{7}{*}{\shortstack[l]{$Robotic_2$\\$LLM_S$\\$VLM_L$}} & Ripple & \mbox{5.11/0.85} & \mbox{6.33/1.05} & \mbox{6.86/0.50} & \mbox{7.80/0.85} & \mbox{8.08/0.45} & \mbox{8.83/0.42} & \mbox{9.44/0.20} \\
 & Sereno UI-First & \mbox{8.14/86.65} & \mbox{8.14/79.72} & \mbox{8.14/71.65} & \mbox{8.14/62.87} & \mbox{8.14/50.80} & \mbox{8.14/12.20} & \mbox{8.14/0.65} \\
 & Sereno Balanced & \mbox{10.55/99.42} & \mbox{10.55/98.92} & \mbox{10.55/96.78} & \mbox{10.55/88.60} & \mbox{10.55/69.32} & \mbox{10.55/21.15} & \mbox{10.55/1.15} \\
 & SiGAMMA & \mbox{6.55/96.43} & \mbox{6.55/83.70} & \mbox{6.55/45.93} & \mbox{6.55/12.35} & \mbox{6.55/0.82} & \mbox{6.55/0.00} & \mbox{6.55/0.00} \\
 & Pause & \mbox{9.06/98.58} & \mbox{9.06/96.23} & \mbox{9.06/94.05} & \mbox{9.06/89.88} & \mbox{9.06/79.88} & \mbox{9.06/10.18} & \mbox{9.06/0.00} \\
 & Cap-$x$ & \mbox{9.21/99.80} & \mbox{8.74/98.80} & \mbox{8.74/94.45} & \mbox{8.74/83.93} & \mbox{8.74/66.88} & \mbox{8.74/4.60} & \mbox{9.21/0.00} \\
 & Unmanaged & \mbox{10.50/100.00} & \mbox{10.50/99.88} & \mbox{10.50/98.65} & \mbox{10.50/91.25} & \mbox{10.50/78.10} & \mbox{10.50/27.15} & \mbox{10.50/2.40} \\
\bottomrule
\end{tabular}
\par\vspace{4pt}
\begin{minipage}{\textwidth}\small At $k=1.5$ for $Robotic_1$ / $LLM_S$ / $VLM_L$, Ripple delivers 7.00 tokens/s at 0.78\% DMR; SiGAMMA delivers 6.16 tokens/s at 21.50\% DMR. The same table shows how throughput and miss rates change across workloads and deadlines.\end{minipage}
\end{table*}

Each 10\,Hz task contributes 1000 jobs per score, and Control contributes
5000. Costmap, KDNN, and SGM use cores 15--18, 11--14, and 7--9; Control
and ORB use 15 and 16. Both engines use cores 0--6. The LLM repeatedly
serves a 37-token prompt with 51 output tokens. VLM periods/deadlines are
5\,s for $VLM_L$ and 10\,s for $VLM_S$. CPU deadlines use the co-run
p99 profile associated with each repetition, with both engines resident but idle. Published baselines use the
same engine flags as \sys{}, including eager LLM execution. Pause and Cap
act at stock forward-pass boundaries; \sys{} uses finer hooks and reacts to
the declared CPU deadline.
The VLM due-time fraction is $\gamma=0.95$.

\para{Baseline configuration.} For Sereno, we tune PI gains separately for
each workload and host using its Ziegler--Nichols procedure: proportional-only
gain rungs are $(1,2,4,\ldots,64)\times10^6$ in parts-per-million units;
the ultimate gain and period come from sustained oscillation or the relay
fallback, and $K_p=0.45K_u$, $K_i=0.54K_u/T_u$. UI-First uses a contention
target of 0.07; Balanced uses 0.15 and a 12-token/s guardrail. The contention
score compares streamed LLM decode-step intervals with their solo median;
the controller pauses only the LLM between steps, without speculative decoding.
For SiGAMMA, a separate process launches a highest-priority spin kernel with
512 threads per block and enough blocks to fill the thread slots on all 48 SMs.
Each CPU job waits for the blocker to become resident before executing; the
blocker remains active while any CPU job is active. For Cap-$x$, each job's
GPU exposure budget is $x$ times its task's solo median duration, with
$x\in\{1/8,1/4,1/2\}$. Exposure accumulates while the job is active and the
engines are commanded to run; when any active job exhausts its budget, the
controller pauses both engines at their next forward-pass boundary.

For the CPU-load comparison, both arms use 8 decode SMs, 48 prefill SMs,
140\,s preparation, and a 200\,s score. Both restore the same saved initial
state and update for 30\,s; the control then freezes CPU adaptation state.
Two 128\,MiB readers on cores 4 and 10 run during 30--80\,s and
130--180\,s, with 500 KDNN offers per interval. Each session learns separately
before the control freezes its adaptation state.

The following experiments use \workloadmix{1}{S}{L} at $k=1.5$ and
140\,s preparation unless specified otherwise.

\para{Sensitivity.}
\label{app:sensitivity}
Table~\ref{tab:sensitivity-current} retains the three settings discussed
in the body. Each changes one parameter from $H=18$\,ms, a 20\,ms lease,
and $W=50$. SM selection runs during preparation, so results include its
selection. Every run meets CPU and VLM targets. Mean throughput changes
by at most 1.7\%; the changes are relative to the current \sys{} control.
\begin{table}[tbp]
\caption{Sensitivity: mean tokens/s per setting.}
\label{tab:sensitivity-current}
\par\vspace{0.1in}
\centering\small
\begin{tabular}{@{}lrr@{}}
\toprule
Setting & Mean tokens/s & Change \\
\midrule
Default & 7.130 & --- \\
$H=12$\,ms & 7.130 & $0.0\%$ \\
Lease 26\,ms & 7.010 & $-1.7\%$ \\
$W=100$ & 7.070 & $-0.8\%$ \\
\bottomrule
\end{tabular}
\end{table}

\para{Transient recovery.}
\label{app:recovery}
A 3\,s GPU copy reaches 187.8--194.4\,GB/s.
Costmap's allowance first returns within 10\% of its pre-copy value after
21.5--23.9\,s, KDNN after 8.5--19.3\,s, and SGM after 14.9--15.2\,s.
Sustained settling is observed for Costmap in a disturbed run within the
post-copy interval. Pooled CPU
misses over the 100\,s scores are 64/73/73 out of 2000 per task
(Costmap/KDNN/SGM), and VLM misses are 4/40.
The disturbed runs fail the CPU and VLM requirements.

\para{Cold start.}
\label{app:cold-start}
Table~\ref{tab:cold-start} compares the first-entry times and initial misses
for starts with and without a CPU profile.
Profiled and no-CPU-prior starts disable certification and its dependent
drift handling in both arms, retaining other calibration. All starts
have zero CPU misses in the first 10\,s (100 jobs per task per run); all
100\,s scores meet CPU and VLM targets.
In a no-prior start, the full pre-score
interval has 41/1470 KDNN and 36/1470 SGM misses, including three dropped
SGM jobs, both above 2\%. %
Profiled/no-prior mean first-entry times
are 57.4/8.4\,s for Costmap, 97.3/60.5\,s for KDNN, and 45.6/63.6\,s for SGM.

\begin{table}[tbp]
\caption{Cold start at $k=1.5$: first entry within 10\% of the final allowance and CPU deadline misses in the first 10\,s, for Costmap / KDNN / SGM.}
\label{tab:cold-start}
\par\vspace{0.1in}
\centering\small
\begin{tabular}{@{}lcc@{}}
\toprule
Start & First entry (s) & Misses, 10\,s \\
\midrule
Profiled & 57.4 / 97.3 / 45.6 & 0 \\
No profile & 8.4 / 60.5 / 63.6 & 0 \\
\bottomrule
\end{tabular}
\par\vspace{3pt}
\begin{minipage}{\linewidth}\small Both starts have zero misses in the first 10\,s. In a no-profile start, the full pre-score interval has 41/1470 KDNN misses and 36/1470 SGM misses.\end{minipage}
\end{table}

\para{One-hour serving.}
\label{app:soak}
The experiment uses a 3600\,s score after 140\,s preparation,
selecting 8 decode SMs. Table~\ref{tab:soak-contract} reports the whole-hour outcomes
and the worst per-task DMR over consecutive 100\,s windows.
All 216 task-windows meet 2\%, and all 1440 VLM frames meet their deadlines.
No throttle flag is active in the 7195 GPU samples.
\begin{table}[tbp]
\caption{One-hour serving at $k=1.5$: CPU deadline miss rates for Costmap / KDNN / SGM and worst task DMR across consecutive 100\,s windows.}
\label{tab:soak-contract}
\par\vspace{0.1in}
\centering\small
\begin{tabular}{@{}lcc@{}}
\toprule
Run & Whole-hour DMR (\%) & Worst window (\%) \\
\midrule
1 & 0.281 / 0.625 / 0.708 & 1.1 \\
2 & 0.253 / 0.719 / 0.572 & 1.1 \\
\bottomrule
\end{tabular}
\end{table}

\end{document}